\documentclass[twocolumn,prl,showpacs,floatfix]{revtex4}
\usepackage{graphicx}
\usepackage{epsfig}
\usepackage{rotating}
\usepackage{amsmath}
\usepackage{amsfonts}
\usepackage{amssymb}
\usepackage{enumerate}
\usepackage{longtable}
\usepackage{dcolumn}
\usepackage{bm}

\begin{document}

\title{Quantum Critical Universality and Singular Corner Entanglement Entropy of Bilayer Heisenberg-Ising model}

\author{Trithep Devakul$^1$ and Rajiv R. P. Singh$^2$}
\affiliation{
    $^1$Department of Physics, Northeastern University, MA 02115, USA \\
$^2$Department of Physics, University of California Davis, CA 95616, USA}

\date{\rm\today}

\begin{abstract}
We consider a bilayer quantum spin model with anisotropic intra-layer exchange couplings.
By varying the anisotropy, the quantum critical phenomena changes from XY to Heisenberg
to Ising universality class, with two, three and one modes respectively becoming gapless simultaneously. 
We use series expansion methods to calculate
the second and third Renyi entanglement entropies when the system is bipartitioned into
two parts. Leading area-law terms and subleading entropies associated with corners are separately
calculated. We find clear evidence that the logarithmic singularity associated with the corners is universal
in each class. Its coefficient along the Ising critical line is in excellent agreement with those obtained previously
for the transverse-field Ising model. Our results provide strong evidence for the idea that the
universal terms in the entanglement entropy provide an approximate measure of the low energy degrees of
freedom in the system.
\end{abstract}


\maketitle

In recent years, the studies of ground state phases of strongly interacting quantum many-body systems 
have been greatly informed and enriched by new ideas from quantum information 
theory\cite{vidal03,verst04,wolf08,rmp_review}. Locality of ground state entanglement propagation
embodied in the ubiquitous `area-law' for entanglement entropy, allows for powerful new variational 
approaches\cite{tensor-network,vidal07,corboz} that have the potential to transform computational materials
science. This has naturally led to great interest in understanding and quantifying the amount of
quantum entanglement present in the true ground states of many-body systems.

\begin{figure}
\begin{center}
\includegraphics[width=7cm]{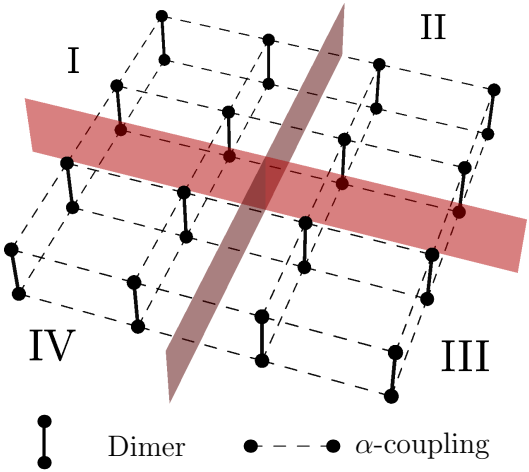}
\caption{\label{fig0} A bilayer of quantum spins can be divided into half-bilayers or quarter-bilayers
by a suitable choice of partitions. These can, in turn, be used to define the `area-law' entanglement entropy
( or the entanglement entropy per unit length of the boundary ) 
and the corner entanglement entropy.
}
\end{center}
\end{figure}

Bipartite entanglement entropy in the ground states of systems with a gap in the 
excitation spectrum is known to obey the area-law\cite{wolf08}, that is, the leading term scales with the `area' 
measuring the boundary between the bipartitioned subsystems. In addition, such systems 
may also contain quantifiable signatures of topological phases in the form of
a long-range entanglement entropy that is unrelated to any boundary\cite{levin_wen,kitaev,isakov,kagome}.
Gapless modes can create additional longer-range entanglement and indeed the study of entanglement properties
can provide novel ways to decipher Goldstone modes associated with spontaneous symmetry breaking\cite{max-grover},
study different quantum critical universality classes\cite{cardy,max}, 
as well as the geometrical properties of fermi surfaces\cite{klich,wolf,haas,swingle,sidel,michelle}.
However, while the studies of universality in entanglement properties of one-dimensional systems
is rather extensive\cite{cardy,peschel}, quantitative studies in higher dimensional lattice models remain few and far
between\cite{melko,series-prl,oitmaa,nlc,nlc-dmrg}.

On approach to the critical point, the entanglement entropies contain both non-universal and universal
singular pieces\cite{rmp_review}. In particular, the corner entropy is expected to have a logarithmic
singularity with a universal coefficient. 
Some universal singular terms may also provide connections to underlying quantum field theories
and the stability of fixed points under renormalization\cite{cardy-rev,grover14,nlc-dmrg}. 
In one dimensional models, the singularities in the
entanglement entropy are known\cite{cardy} to be related to the conformal anomaly $c$, and thus, naturally provide
a connection to Zamalodchikov's c-theorem\cite{zamolodchikov} that posits that under renormalization systems flow towards
smaller $c$-values. There have been many recent efforts to generalize Zamalodchikov's c-theorem
to higher dimensional systems and the entanglement entropy is central to such efforts\cite{huerta}. In a very general sense,
one expects the singularities in the entanglement entropy to provide a measure of low lying fluctuations or a count of degrees
of freedom in the underlying continuum theory\cite{cardy-rev,grover14}. Quantum field theory calculations are
difficult and hence numerical approaches must play an essential role.

Here, we consider a bilayer consisting of two-planes of square-lattice of spins (see Fig.~\ref{fig0}),
with an anisotropic quantum Heisenberg model with Hamiltonian,
\begin{eqnarray}
{\cal H}&=& \sum_{\langle i,j\rangle} \left(S_i^x S_j^x +  S_i^y S_j^y + S_i^z S_j^z\right)\nonumber \\
        &+& \alpha \sum_{\langle\langle i,k\rangle\rangle} \left(S_i^x S_k^x  + S_i^y S_k^y + \lambda S_i^z S_k^z\right),
\end{eqnarray}
where the first sum runs over pairs of spins between the two layers, while the
second sum runs over the nearest-neighbor spins, within each layer. For $\alpha=0$,
this model is in the singlet phase, where each pair of inter-planar spins form a dimer.
When $\alpha$ is large, the system goes from an XY order at small
$\lambda$ to Ising order at large $\lambda$. Right at $\lambda=1$, full Heisenberg symmetry
is realized and we have the well studied bilayer quantum Heisenberg model\cite{nlc-dmrg,heisenberg-qmc,weihong,chakravarty,wessel}. In the singlet
phase there are three single-particle excitations corresponding to $S^z=+1$, $0$ and $-1$
respectively. As is evident from the symmetries of the model for $\lambda<1$, the gap to $S^z=\pm 1$ 
modes closes first leading to $XY$ order. For $\lambda>1$, the gap to $S^z=0$ mode closes first
leading to Ising order. At $\lambda=1$, all three modes remain exactly degenerate and the
system has higher symmetry and Heisenberg universality. Thus, this model provides an elegant way
to study the quantum-critical behavior and different universality classes and their crossovers in a two-dimensional quantum system.

To carry out our series expansions, we consider one value of the anisotropy $\lambda$ at a time.  
We expand around the dimerized phase at $\alpha=0$.
To determine the critical coupling $\alpha_c$ at each $\lambda$, we calculate series for the staggered susceptibility and excitation gap.  
For $\lambda < 1$, we consider the staggered susceptibility to a field in the $x$-direction, and the $S^z = +1$ excitation gap.  
For $\lambda \ge 1$, we instead consider the staggered susceptibility to a field in the $z$-direction, and the $S^z = 0$ excitation gap.

We then calculate series expansions for the `area-law' terms for the 2nd and 3rd Renyi 
entanglement entropies at various values of $\lambda$.
To do this, we consider an $L\times L\times 2$ lattice with free boundary conditions. We bipartition the bilayer of spins
into two equal size half-bilayers by one of the two dividers shown in Fig.~\ref{fig0}.
The left half-bilayer (regions I and IV)
can be denoted A and the right half-bilayer (regions II and III) can be denoted B.
The $n$th Renyi entropy is defined as
\begin{eqnarray}
    S_n(A) &=& \frac{1}{1-n}\ln \text{Tr}\left( \hat{\rho}^n_A \right),
\end{eqnarray}
where $\hat{\rho}_A = \text{Tr}_B \left|{\Psi}\right\rangle \left\langle{\Psi}\right|$ is the reduced density matrix for subsystem $A$.
The length of the interface between A and B is given by $L$. We define the Renyi entanglement
entropy per unit length of the interface, in the thermodynamic limit, as
\begin{equation}
s_n=\lim_{L\to\infty} S_n/L.
\end{equation}
The quantities, $s_n$ can be calculated, as a power series expansion in $\lambda$, by the linked cluster method\cite{book,series-reviews,oitmaa}.

The entropy due to the presence of a 90$^\circ$ corner in the boundary for the $n$th Renyi entropy, $c_n$, 
can also be isolated in our series expansions 
\cite{oitmaa,nlc-dmrg}. 
This is done by calculating the entanglement entropy upon subdivision of the system by two perpendicular partitions 
(see Fig.~\ref{fig0}), either into a quarter-bilayer (and its complement of 3-quarter bilayers) or into two half bilayers. 
The contributions from the linear boundaries can be cancelled out, by subtracting the entropies when the subsystem A
is a half-bilayer from the case when it is a quarter bilayer \cite{oitmaa}. This allows us to generate a 
separate series expansion for the corner entropy $c_n$.

All series coefficients are calculated up to $\alpha^{11}$.
A single calculation of this order can be completed within a day on a moderately powerful personal computer.

\begin{figure}
\begin{center}
\includegraphics[width=7cm,angle=270]{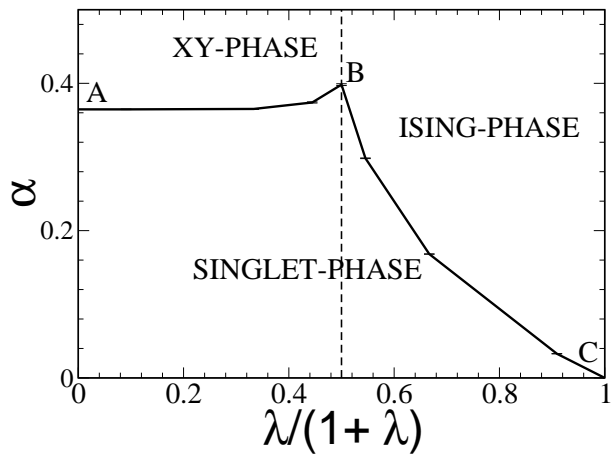}
\caption{\label{fig1} The phase diagram of the model with the Singlet phase at small $\alpha$ and the ordered
XY and Ising phases at large $\alpha$. The dashed line represents the system with full Heisenberg symmetry. The
critical point $B$ has $O(3)$ universality class and separates  $O(2)$ universality class along $AB$
from Ising universality class along $BC$.
}
\end{center}
\end{figure}
\begin{figure}
\begin{center}
\includegraphics[width=7cm,angle=270]{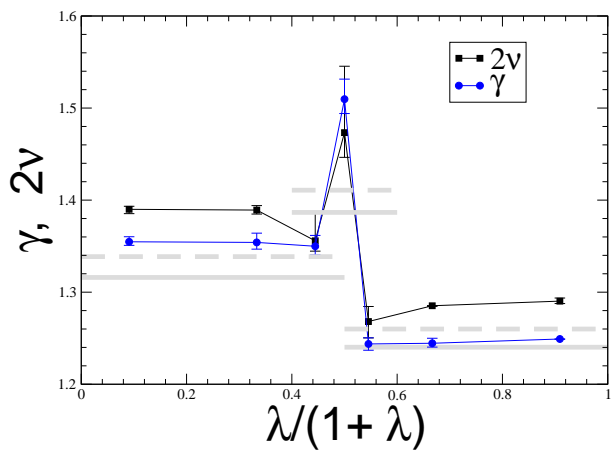}
\caption{\label{fig2} Critical exponents $\gamma$ and $\nu$ are shown for different values of the anisotropy parameter $\lambda$.
Our results are consistent with constant values for the exponents in the XY and Ising regimes, with a larger exponent at
the special Heisenberg symmetry case. The results from epsilon expansions\cite{zinn-justin} are shown as solid lines
for $\gamma$ and dashed lines for $2\nu$ (On the right for Ising, on the left for XY and in the middle for classical
Heisenberg universality class).
}
\end{center}
\end{figure}

The series are analyzed using Pade and differential approximants\cite{book,series-reviews}. First we analyze the
series for the susceptibility and gap series. These lead to the determination of the phase-boundary shown in
Figure \ref{fig1}. The critical couplings $\alpha_c$ obtained from the susceptibility and gap series are consistent
with each other. The critical exponents $\gamma$ for the susceptibility divergence and $\nu$ for the gap closing (taking $z=1$)
are calculated and shown in Figure \ref{fig2}. Here, and throughout this paper, the error bars shown represent a spread in the different
approximants and are not true statistical uncertainties. The results from epsilon expansion calculations, taken from the
literature\cite{zinn-justin} are also shown.
On the Ising side the exponents are within 1-2 percent of accepted values for the 3-dimensional 
Ising universality class, while on the XY
side they are within 2-3 percent of accepted values for the 3-dimensional O(2) universality class\cite{zinn-justin}. 
Only at the Heisenberg point ($\lambda=1$), 
the deviations are larger (approximately 6-7 percent for the susceptibility exponent $\gamma$). These are largely correlated with
uncertainties in $\alpha_c$, which varies sharply near the Heisenberg point. For example, the susceptibility series leads to estimates of $\alpha_c=0.3982\pm 0.0008$,
with $\gamma=1.51\pm 0.02$. If we bias approximants to much more accurate values of the critical point from Quantum Monte
Carlo simulations\cite{heisenberg-qmc} $\alpha_c=0.39651$, it leads to estimates for exponent of $\gamma=1.43\pm 0.02$,
which are much closer to accepted values for the 3-dimensional classical O(3) universality class.

\begin{figure}
\begin{center}
\includegraphics[width=7cm,angle=270]{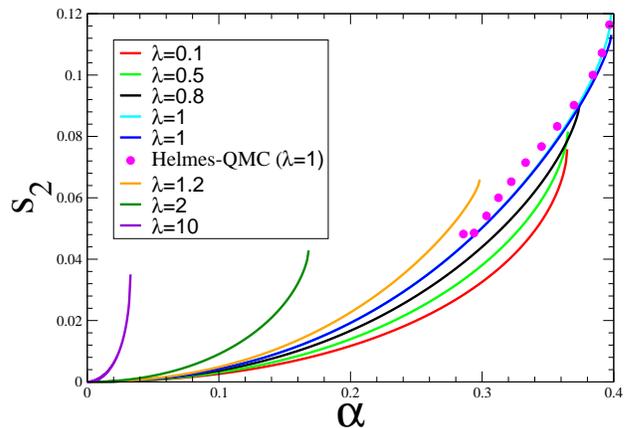}
\caption{\label{fig3} Plots of the `Area-law' or entanglement entropy per unit boundary length
for the second Renyi entropy $S_2$. Quantum Monte Carlo data from Helmes et al are shown by filled circles.
For $\lambda=1$ two approximants are shown for $s_2/\lambda^2$ series: a [4,4] Pade approximant and a [$T=0$, $M=3$, $L=5$] biassed 
first-order Integrated Differential Approximant \cite{book}.
}
\end{center}
\end{figure}

\begin{figure}
\begin{center}
\includegraphics[width=7cm,angle=270]{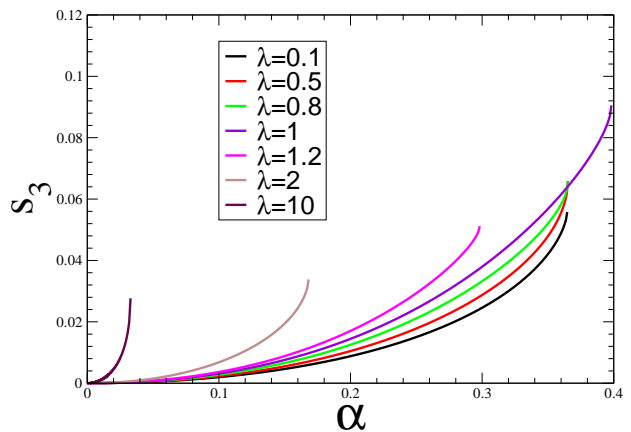}
\caption{\label{fig4}
Plots of the `Area-law' or entanglement entropy per unit boundary length
for the third Renyi entropy $S_3$.
}
\end{center}
\end{figure}

The area-law term in the entanglement entropy is known to have a weak singularity at the critical coupling. We simply
used biased differential approximants, with the critical points $\alpha_c$ estimated from the susceptibility
series, to obtain their values up to the critical coupling. The plots for $s_2$ and $s_3$ are shown in Figures \ref{fig3}
and \ref{fig4} respectively. Our results should be highly accurate except possibly when one is within a few percent
of the critical coupling. Two different approximants are plotted for $s_2$ for $\lambda=1$ to show the expected deviations.
The difference is barely visible in the plots.
For the second Renyi entropy a comparison with recent Quantum Monte Carlo study\cite{wessel} is also made. 
The overall agreement is very good. Results from series expansion should be much more accurate away from
the critical point.

The quantity of primary interest in our study is the corner entropy and its singular behavior.
On approach to the critical coupling, we expect the corner entropy to behave as
\begin{equation}
c_n = a_n \ln{\xi}=-\nu \ a_n \ln{(\alpha_c-\alpha)}.
\end{equation}
Here $a_n$ should be a universal constant equal to the coefficient in the
logarithmic size dependence $a_n \ln{L}$ of the entanglement entropy
for a square region of linear dimension $L$ when the system is
right at the critical coupling\cite{max,nlc-dmrg}.

\begin{figure}
\begin{center}
\includegraphics[width=7cm,angle=270]{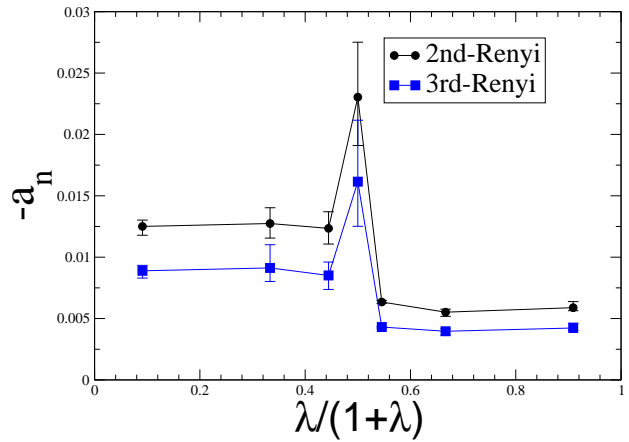}
\caption{\label{fig5}
Coefficient $-a_n$ of the logarithmic singularity associated with a corner in the boundary.
These results show that the Ising and XY universality classes have constant coefficients.
At Heisenberg symmetry the coefficient is largest. Our results are consistent with a rough
proportionality for the log coefficients with the number of gapless modes in the system.
}
\end{center}
\end{figure}

To analyze the corner entropy, we first take a derivative of the series, which converts a logarithmic
singularity into a simple pole. We then study this by a simple Pade approximant biasing the critical
point to that obtained from the analysis of the susceptibility series. The estimated $a_n$ values are shown
in Figure \ref{fig5}. In the Ising limit, our results are in excellent agreement with previous series
expansions\cite{oitmaa} and Numerical Linked Cluster (NLC) calculations\cite{nlc} for the transverse-field
Ising model. Very near the Heisenberg
point, one expects greater uncertainty in the values due to crossover effects. Hence, we consider the
values at $\lambda=2$ and $\lambda=10$. At $\lambda=2$ we estimate $a_2=-0.0055\pm 0.0003$ while at
$\lambda=10$ we estimate $a_2=-0.0059\pm 0.0005$. These values agree well with the transverse field
Ising model values of $-0.0055\pm 0.0005$ from series expansions\cite{oitmaa,fn} and $-0.0053$ from the
NLC calculations\cite{nlc}. Quantum Monte Carlo estimates so far have\cite{roscilde} much larger error bars $-0.0075\pm 0.0025$,
but are also consistent with these estimates.
For the third Renyi entropy, we estimate $a_3=-0.0040\pm 0.0003$ at $\lambda=2$ and $a_3=-0.0042\pm 0.0004$ at
$\lambda=10$. These values are also in excellent agreement with the transverse-field Ising model values calculated
by NLC. These results point strongly towards the universality of these coefficients
in the Ising phase of this model and also between this model and the transverse-field Ising model.

We are not aware of any previous calculations of log singularities for the XY universality class.
Our results lead to estimates of $a_2=-0.0125\pm 0.0006$ at $\lambda=0.2$ and $a_2=-0.0127\pm 0.0013$
at $\lambda=0.5$. Similar results for the third Renyi entropy are $a_3=-0.0089\pm 0.0006$ at $\lambda=0.2$
and $a_3=-0.0091\pm 0.0020$ at $\lambda=0.5$. These results clearly point towards a universal value
for this universality class also.

At the Heisenberg point ($\lambda=1$) our results have greater uncertainty. 
In the Fig.~\ref{fig5}, the $a_n$ values
are shown for the critical point biased at $\alpha_c=0.3982$ (estimate from susceptibility series), which leads to $a_2=-0.023\pm 0.005$
and $a_3=-0.016\pm 0.005$. If we bias the critical point to $\alpha_c=0.3965107$ known
more accurately from previous studies\cite{heisenberg-qmc}, we get 
$a_2=-0.022\pm 0.005$ 
and $a_3=-0.016\pm 0.005$. Corner entanglement entropy of this
model has been studied before using NLC\cite{nlc-dmrg} and QMC\cite{wessel}. The NLC values were found to be very close to
$3$ times the transverse-field Ising model values for all Renyi indices. Also, the QMC fits gave $a_2=-0.016\pm 0.001$.
Our numbers are somewhat larger than these but still in fair agreement with them.
Overall, it is clear that the results imply a rough proportionality between the singular coefficients and the
number of soft modes in the system that become gapless at the critical point.

In conclusion, in this paper we have studied a bilayer XXZ model that allows us to tune between
Ising, Heisenberg and XY universality classes. We have calculated the `area-law' terms and corner entanglement
Renyi entropies using series expansion. We presented strong evidence that the corner entanglement has
a logarithmic singularity, which takes universal values in different universality classes. Furthermore,
this universal coefficient is roughly proportional to the number of soft or gapless modes in the system.
Within these models, its magnitude decreasees monotonically from less stable to more stable fixed points.
However, the question of whether it is truly monotonic under renormalization
and can serve the purpose of organizing stability of higher dimensional fixed points
of quantum statistical models,
analogous to the central charge $c$ in one-dimensional models, deserves further attention.

\begin{acknowledgements}
We would like to thank Tarun Grover and Roger Melko for many enlightening communications.
This work is supported in part by NSF grant number  DMR-1306048.
\end{acknowledgements}


\begin{thebibliography}{2}

\bibitem{vidal03} G. Vidal, J.I. Latorre, E. Rico,and A. Kitaev, Phys. Rev. Lett. 90, 227902 (2003).

\bibitem{verst04} F. Verstraete, M. Popp, and J.I. Cirac, Phys. Rev. Lett. 92, 027901 (2004).

\bibitem{wolf08}  M. M. Wolf, F. Verstraete, M. B. Hastings, and J. I. Cirac
Phys. Rev. Lett. 100, 070502 (2008).

\bibitem{rmp_review} J. Eisert, M. Cramer and M.B. Plenio, Rev. Mod. Phys. {\bf 82}, 277 (2010).

\bibitem{tensor-network} J. Jordan, R. Orús, G. Vidal, F. Verstraete, and J. I. Cirac
Phys. Rev. Lett. 101, 250602 (2008).

\bibitem{vidal07} G. Vidal, Phys. Rev. Lett. 99, 220405 (2007).

\bibitem{corboz} Philippe Corboz, T. M. Rice and Matthias Troyer, cond-mat/ arXiv:1402.2859.

\bibitem{levin_wen} M. Levin and X.G. Wen, Phys. Rev. Lett. 96, 110405 (2006).

\bibitem{kitaev} A. Kitaev and J. Preskill,
Phys. Rev. Lett. 96, 110404 (2006).

\bibitem{isakov} S. V. Isakov, R. G. Melko, M. B. Hastings, Science {\bf 335}, 193 (2012).

\bibitem{kagome} S. Yan, D. A. Huse and S. R. White, Science 332, 1173 (2011); 
S. Depenbrock, I. P. McCulloch and U. Schollwock, Phys. Rev. Lett. 109, 067201 (2012);
H. C. Jiang, Z. H. Wang and L. Balents, Nat. Phys. 8, 902 (2012).

\bibitem{max-grover} Max A. Metlitski, Tarun Grover,  arXiv:1112.5166.

\bibitem{cardy}  P. Calabrese and J. Cardy, J. Stat. Mech: Theor. Exp. P06002 (2004).

\bibitem{max}  M. A. Metlitski, C. A. Fuertes, and S. Sachdev
Phys. Rev. B 80, 115122 (2009).

\bibitem{klich} D. Gioev and I. Klich, Phys. Rev. Lett. 96, 100503 (2006).

\bibitem{wolf} M. Wolf, Phys. Rev. Lett. 96, 010404 (2006).

\bibitem{haas} W. Li, L. Ding, R. Yu, T. Roscilde, and S. Haas,
Phys. Rev. B 74, 073103 (2006).

\bibitem{swingle} B. Swingle, Phys. Rev. Lett. 105, 050502 (2010).

\bibitem{sidel} W. Ding, A. Seidel, and K. Yang
Phys. Rev. X 2, 011012 (2012).

\bibitem{michelle} Michelle Storms and Rajiv R. P. Singh,
Phys. Rev. E 89, 012125 (2014).

\bibitem{peschel} I. Peschel and V. Eisler, J. Phys. A-Math. and Theor. 42, 504003 (2009);
V. Eisler and I. Peschel, J. Stat. Mech-Theor. and Exp. P06005 (2007).

\bibitem{melko} M.~B. Hastings, I.~Gonz\'alez, A.~B. Kallin, R.~G. Melko,  Phys. Rev. Lett. {\bf 104}, 157201 (2010).

\bibitem{series-prl} R. R. P. Singh, M. B. Hastings, A. B. Kallin and R. G. Melko,
Phys. Rev. Lett. 106, 135701 (2011).  A. B. Kallin, M. B. Hastings, R. G. Melko and R. R. P. Singh,
Phys. Rev. B 84, 165134 (2011).

\bibitem{oitmaa} R. R. P. Singh, R. G. Melko, and J. Oitmaa,
Phys. Rev. B 86, 075106 (2012).

\bibitem{nlc} A. B. Kallin, K. Hyatt, R. R. P. Singh, and R. G. Melko
Phys. Rev. Lett. 110, 135702 (2013).

\bibitem{nlc-dmrg} Ann B. Kallin, E. M. Stoudenmire, Paul Fendley, Rajiv R. P. Singh, Roger G. Melko,
cond-mat arXiv:1401.3504.

\bibitem{cardy-rev} J. Cardy, arXiv:1008.2331, J. Stat. Mech. 1010:P10004 (2010).

\bibitem{grover14} T. Grover, Phys. Rev. Lett. 112, 151601 (2014).

\bibitem{zamolodchikov} A. B. Zamolodchikov, JETP Lett. 43, 731 (1986).

\bibitem{huerta} H. Casini and M. Huerta, Phys. Rev. D 85, 125016 (2012).

\bibitem{heisenberg-qmc} L. Wang, K. S. D. Beach and A. W. Sandvik, Phys. Rev. B 73, 014431 (2006).

\bibitem{weihong} Z. Weihong, Phys. Rev. B 55, 12267 (1997); C. J. Hamer, J. Oitmaa and Z. Weihong,
Phys. Rev. B 85, 104432 (2012).

\bibitem{chakravarty} L. Yin, M. Troyer and S. Chakravarty, Europhys. Lett. 52, 559 (1998).

\bibitem{wessel} Johannes Helmes, Stefan Wessel,  arXiv:1403.7395.


\bibitem{book} J. Oitmaa, C. Hamer and W. Zheng, {\it Series Expansion
Methods for strongly interacting lattice models} (Cambridge University
Press, 2006).

\bibitem{series-reviews}  M. P. Gelfand and R. R. P. Singh, Adv. Phys. {\bf 49}, 93(2000);
M. P. Gelfand, R. R. P. Singh and D. A. Huse, J. Stat. Phys. 59, 1093 (1990).

\bibitem{zinn-justin} J. C. Le Guillou and J. Zinn Justin, Phys. Rev. Lett. 39, 95 (1977).

\bibitem{fn} The results presented for transverse field Ising model in Ref.~24, is
off by a factor of 2. The entire series in the supplement of the paper is for 2 times a corner singularity,
and so is the quoted log coefficient. We have explicitly checked this by recalculating the series.

\bibitem{roscilde} S. Humeniuk and T. Roscilde, Phys. Rev. B 86, 235116 (2012).


\end{thebibliography}

\end{document}